\documentclass[useAMS,usenatbib]{mn2e}
\usepackage[dvips]{graphicx}
\usepackage{hyperref}
\usepackage{amsmath}
\usepackage{amssymb}
\usepackage{caption}
\include{aas_macros}

\newcommand{\teff}{$T_{\rm eff}$}

\newcommand{\mdot}{$\dot{M}$}

\newcommand{\rk}{$R_{\rm K}$}
\newcommand{\ra}{$R_{\rm A}$}

\newcommand{\beq}{\begin{equation}}
\newcommand{\eeq}{\end{equation}}
\newcommand{\beqa}{\begin{eqnarray}}
\newcommand{\eeqa}{\end{eqnarray}}

\newcommand{\mnras}{$MNRAS$}
\newcommand{\apj}{$ApJ$}
\newcommand{\apjl}{$ApJ$}

\newcommand{\aap}{$A\&A$}

\newcommand{\apjs}{$ApJS$}

\newcommand{\jgr}{$JGR$}

\newcommand{\icarus}{Icarus}
\newcommand{\emf}{\mathcal{E}}
\newcommand{\res}{\mathcal{R}}

\title[Centrifugal Breakout and Gyrosynchroton Emission]{Centrifugal breakout reconnection as the electron acceleration mechanism powering the radio magnetospheres of early-type stars}
\author[S.\ P.\ Owocki et al.]{S.\ P.\ Owocki$^{1}$\thanks{E-mail:
owocki@udel.edu},
M.\ E.\ Shultz$^1$,
A.\ ud-Doula$^2$,
P.\ Chandra$^3$,
B.\ Das$^3$,
P.\ Leto$^{4}$
\\
$^1$Department of Physics and Astronomy, University of Delaware, 217 Sharp Lab, Newark, Delaware, 19716, USA\\
$^2$Department of Physics, Penn State Scranton, Dunmore, PA 18512, USA\\
$^3$National Centre for Radio Astrophysics, Tata Institute of Fundamental Research, Pune University Campus, Pune-411007, India\\
$^4$INAF - Osservatorio Astrofisico di Catania, Via S. Sofia 78, 95123 Catania, Italy\\
}
\begin{document}

\date{}

\pagerange{\pageref{firstpage}--\pageref{lastpage}} \pubyear{2002}

\maketitle

\label{firstpage}

\begin{abstract}
Magnetic B-stars often exhibit circularly polarized radio emission thought to arise from gyrosynchrotron emission by energetic electrons trapped in the circumstellar magnetosphere. Recent empirical analyses show that the onset and strength of the observed radio emission scale with both  the magnetic field strength and the stellar rotation rate. This challenges the existing paradigm that the energetic electrons are accelerated in the current sheet between opposite-polarity field lines in the outer regions of magnetised stellar winds, which includes no role for stellar rotation. Building on recent success in explaining a similar rotation-field dependence of H$\alpha$ line emission in terms of a model in which magnetospheric density is regulated by centrifugal breakout (CBO), we examine here the potential role of the associated CBO-driven magnetic reconnection in accelerating the electrons that  emit the observed gyrosynchrotron radio. We show in particular that the theoretical scalings for energy production by CBO reconnection match well the empirical trends for observed radio luminosity, with a suitably small, nearly constant conversion efficiency $\epsilon \approx 10^{-8}$. We summarize the distinct advantages of our CBO scalings over previous associations with an electromotive force, and discuss the potential implications of CBO processes for X-rays and other observed characteristics of rotating magnetic B-stars with   centrifugal magnetospheres.
\end{abstract}


\section{Introduction}
Hot luminous, massive stars of spectral type O and B have dense, high-speed, radiatively driven stellar winds
\citep{cak1975}.
In the subset ($\sim$10\%; \cite{2017MNRAS.465.2432G,2019MNRAS.483.2300S}) of massive stars with strong ($>\,100$\,G; \cite{2007AA...475.1053A,2019MNRAS.482.3950S}), globally ordered (often significantly dipolar; \citet{2019A&A...621A..47K}) magnetic fields, the trapping of this wind outflow by closed magnetic loops leads to the formation of a circumstellar {\em magnetosphere} \citep{petit2013}.
Because of the angular momentum loss associated with 
their relatively strong, magnetised wind 
\citep{ud2009}, magnetic O-type stars are typically
slow rotators,  with trapped wind material falling back on a dynamical timescale, giving what's known as a ``dynamical magnetosphere" (DM).

But in magnetic B-type stars, the relatively weak stellar winds imply longer spin-down times, and so a significant fraction that still retain a moderately rapid rotation;
for cases in which the associated Keplerian co-rotation radius $R_{\rm K}$ lies within the Alfv\'{e}n radius $R_{\rm A}$ that characterises the maximum height of closed loops, the rotational 
support leads to formation of a ``{\em centrifugal magnetosphere}'' (CM), 
wherein the much longer confinement time allows material to build up to a sufficiently high density to give rise to distinct emission in H$\alpha$ and other hydrogen lines \citep{lb1978}.
A recent combination of empirical \citep{2020MNRAS.499.5379S} and theoretical \citep{2020MNRAS.499.5366O} analyses showed that both the   onset and strength of such Balmer-$\alpha$ emission is well explained by a {\em centrifugal breakout} (CBO) model, 
wherein the density distribution of material within the CM is regulated to be near the critical level that can be contained by magnetic tension
\citep{ud2008}.
The upshot is that such hydrogen emission arises only in magnetic stars with {\em both} strong magnetic confinement and moderately rapid stellar rotation.

\begin{figure*}
   \centering
   \includegraphics[width=0.95\textwidth]{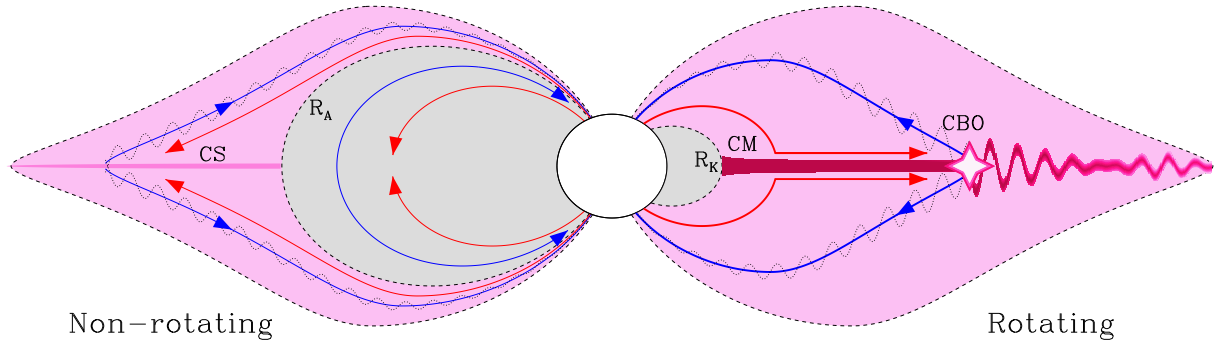} 
      \caption[]{
 Schematic to contrast the previous \citet{2004A&A...418..593T} current-sheet (CS) model for electron acceleration and radio emission (left)
 with our proposed {\em centrifugal breakout} (CBO) model for  electron acceleration from CBO-driven magnetic reconnection events (right).
 Pink-shaded regions indicate magnetic field lines contributing plasma to the electron acceleration, while
 grey shading indicates regions isolated from the locus of electron acceleration.
 A key distinction regards the lack of a dynamical role for rotation in the CS model, in contrast to the inferred empirical dependence on rotation,
 which is well matched in the CBO model.
 Figure adopted from Figure 8 of Paper I, which contains further details on the schematic.
}
 \label{cartoon}
\end{figure*}

Another distinctive observational characteristic of many such magnetic B-stars is their  non-thermal, circularly polarized {\em radio} emission, thought to arise from gyrosynchrotron emission by energetic electrons trapped within closed magnetic loops.
An initially favoured model by  \citet{2004A&A...418..593T} proposed that these electrons could be accelerated in the current sheet  between field lines of opposite polarity that have been stretched outward by the stellar wind,  as illustrated in the left panel of Figure \ref{cartoon}.
But a recent empirical analysis by \citet{2021MNRAS.507.1979L} has shown that the observed radio emission has a clear dependence on stellar rotation, providing strong evidence against this 
current-sheet model, which includes no  role for rotation.
Instead \citet{2021MNRAS.507.1979L} noted that their fits to the radio luminosity scale in proportion to a quantity that has the physical dimension of an electromotive force (EMF), which they speculated may be suggestive of an underlying mechanism.
Indeed,  the EMF is invoked 
\citep{2001JGR...106.8101H} 
to model auroral emission from the interaction of high-energy magnetospheric particles with planetary atmospheres.
However, such thermal atmospheric auroral dissipation of EMF-accelerated particles in the magnetosphere cannot explain the polarized radio emission that likely arises from gyrosynchrotron processes in the highly conductive magnetosphere itself.

The alternative theoretical scalings explored here were motivated by a more recent companion empirical analysis by 
\citet[][hereafter Paper I]{shultz2021},
which confirms the basic results of \citet{2021MNRAS.507.1979L}, but within a significantly extended sample that allows further exploration of potential empirical trends and scalings.
In particular, we show below (section \ref{sec:cbomc}) that these empirical scalings for nonthermal radio emission can be well fit by models grounded in the same CBO paradigm that has been so successful for H$\alpha$ emission.
Specifically, it is now the {\it magnetic reconnection} associated with CBO events that provides the nonthermal acceleration of electrons, which then follow the standard picture of gyrosynchrotron emission of observed circularly polarized radio.

Figure \ref{cartoon} graphically illustrates the key distinctions between this new CBO paradigm (right) from the previous current-sheet-acceleration model (left) proposed by \citet{2004A&A...418..593T}.

To lay the groundwork for derivation in section \ref{sec:cbomc} of these scalings for  radio emission from CBO-driven reconnection, the next section  (\ref{sec:scaling_law}) reviews the basic CM model and the previous application of the CBO paradigm to H$\alpha$.
In section \ref{sec:disc} we contrast our results with the EMF-based picture, and discuss the potential further application of the CBO paradigm, including for modeling the stronger X-ray emission of CM vs. DM stars \citep{2014ApJS..215...10N}.
We conclude (section \ref{sec:sum}) with a brief summary and outlook for future work.

\section{Background}\label{sec:scaling_law}

\subsection{Dynamical vs.\ Centrifugal magnetospheres}

For a magnetic hot-star with stellar wind mass loss rate ${\dot M}$ and terminal wind speed $v_\infty$, MHD simulations \citep{ud2002} show that the channeling and trapping of the stellar wind can  be characterised by a dimensionless wind-magnetic-confinement parameter,
\beq
\eta_\ast \equiv \frac{B_{\rm eq}^2R_\ast^2}{\dot{M}v_\infty}
\, ,
\label{eq:etastar}
\eeq
where $B_{\rm eq}$ is the surface field strength at the magnetic equator and $R_\ast$ is the stellar radius.
This characterises the ratio of magnetic energy to wind kinetic energy.
The radial extent of closed magnetic loops can be characterised by the  Alfv\'en radius, which for  an initially dipolar field with strong-confinement  scales as
\beq
\frac{R_{\rm A}}{R_\ast} \approx \eta_\ast^{1/4}  ~~ ; ~~ \eta_\ast \gg 1
\, ,
\label{eq:RAdip}
\eeq

Simulations of cases with rotation-aligned dipoles \citep{ud2008} showed further that the dynamical effect of rotation can be similarly characterised by a dimensionless parameter, now given by the ratio of the star's equatorial rotation speed to the near-star orbital speed,
\begin{equation}
W = \frac{v_{\rm rot}}{v_{\rm orb}} = \frac{2\pi R_\ast}{P_{\rm rot}}\left(\frac{GM_\ast}{R_\ast}\right)^{-1/2}
\, ,
\label{eq:wrot}
\end{equation}
with $M_\ast$ and $P_{\rm rot}$ the stellar mass and rotation period, and $G$ the gravitation constant.
For magnetically trapped material that is forced to co-rotate with the underlying star, centrifugal forces balance gravity in the common equator at the Kepler co-rotation radius, given by
\beq
\frac{R_{\rm K} }{R_\ast} = W^{-2/3}
\, ,
\label{eq:RK}
\eeq
For slowly rotating stars with $R_{\rm K} > R_{\rm A}$, rotation has little dynamical effect, and so wind material trapped in closed magnetic loops below $R_{\rm A}$ simply falls back to the star on a dynamical timescale, giving then a {\em dynamical magnetosphere} (DM).

In contrast, for stars with both moderately rapid rotation ($W \lesssim 1$) and strong confinement ($\eta_\ast \gg 1$), one finds
$R_{\rm K} < R_{\rm A}$. In the region $R_{\rm K} < r < R_{\rm A}$ magnetic tension still confines material while the centrifugal force prevents gravitational fallback, thus allowing material build-up into a much denser {\em centrifugal magnetosphere} (CM).

\subsection{Centrifugal breakout and H$\alpha$ emission}

As first analyzed in the appendices of \citet{town2005c}, this CM mass buildup is limited to a critical surface density for which the finite magnetic tension can still confine the material against the outward centrifugal acceleration.
For mass buildup beyond this critical density, the magnetic field lines become stretched outward by the centrifugal force, leading eventually to 
{\em centrifugal breakout} (CBO) events.
Through analysis of 2D MHD simulations by \citet{ud2008}, \citet{2020MNRAS.499.5366O} showed that the resulting global surface density  scales as
\beq
\sigma (r) \approx \sigma_{\rm K} \,
\left ( \frac{r}{R_{\rm K}}
\right )^{-6} 
 ~~ ; ~~ r > R_{\rm K}
\, ,
\label{eq:sigr}
\eeq
where the characteristic surface density at the Kepler radius scales with the magnetic field strength and gravitational acceleration there,
\beq
\sigma_{\rm K} \approx 0.3 \, \frac{B_{\rm K}^2}{4 \pi g_{\rm K}}
\, .
\label{eq:sigK}
\eeq

A key feature of this CBO-regulated density is that it is entirely {\em independent} of the stellar wind mass loss rate ${\dot M}$ that controls the CM mass buildup.
This helps explain the initially unexpected empirical finding by \citet{2020MNRAS.499.5379S} that the onset and strength of observed H$\alpha$ emission from magnetic B-stars is largely independent of the stellar luminosity, which plays a key role in setting the mass loss rate of the radiatively driven stellar wind.

Motivated by this key result,
\citet{2020MNRAS.499.5366O} examined the theoretical implications of this CBO-limited density scaling for such H$\alpha$ emission,
showing that it can simultaneously explain the onset of emission, the increase of emission strength with increasing magnetic field strength and decreasing rotation period, and the morphology of emission line profiles \citep{2020MNRAS.499.5379S,2020MNRAS.499.5366O}. 
As initially suggested by \cite{town2005c}, the breakout density at \rk~is set by $B_{\rm K}$, and is independent of \mdot; 
precisely this dependence on $B_{\rm K}$, and lack of sensitivity to \mdot, was found by \cite{2020MNRAS.499.5379S} for both emission onset and emission strength scaling. \cite{2020MNRAS.499.5366O} found an expression for the strength $B_{\rm K1}$ necessary for the density at \rk~to produce an optical depth of unity in the H$\alpha$ line, and showed that the threshold $B_{\rm K}/B_{\rm K1}$ neatly divides stars with and without H$\alpha$ emission. Two-dimensional MHD simulations of CBO by \cite{ud2006,ud2008} yielded a radial density gradient associated with the CBO mechanism, which in conjunction with the density at \rk~set by $B_{\rm K}$ can be used to predict the optically thick area and, hence, the scaling of emission strength \citep{2020MNRAS.499.5366O}. Finally, a characteristic emission line profile morphology, common across all H$\alpha$-bright CM host stars, was reported by \cite{2020MNRAS.499.5379S} and shown by \cite{2020MNRAS.499.5366O} to be a straightforward consequence of a co-rotating optically thick inner disk transitioning to optically transluscent in the outermost region. 

A crucial subtlety that deserves emphasis is that, in contrast to expectations from 2D MHD simulations that CBO should manifest as catastrophic ejection events accompanied by large-scale reorganization of the magnetosphere \citep{ud2006,ud2008}, which has indeed never been observed in the densest inner regions \citep{town2013,2020MNRAS.499.5379S}, the H$\alpha$ analysis performed by \cite{2020MNRAS.499.5379S} instead indicates that the magnetosphere must be continuously maintained at breakout density, with CBO occuring more or less continuously on small spatial scales. However, it is worth noting that the `giant electron-cyclotron maser (ECM) pulse' observed by 
\cite{das2021}
may have been the signature of a large-scale breakout occuring in magnetospheric regions in which the density is too low to be probed by H$\alpha$ or photometry.

H$\alpha$ emission and gyrosynchrotron emission occur in the same part of the rotation-magnetic confinement diagram (see Fig.\ 3 in Paper I), and H$\alpha$ emission EW and radio luminosity are closely correlated (see Fig.\ 6 in Paper I). Since H$\alpha$ emission is regulated by CBO, this suggests that the same may be true of gyrosynchrotron emission. In the following we develop a theoretical basis for this connection, which we then compare to the empirical regression analyses and measured radio luminosities. 

\section{CBO-driven Magnetic Reconnection}
\label{sec:cbomc}

\subsection{Rotational spindown}


The rotational energy of a star with moment of inertia $I$ and rotational frequency $\Omega$ is given by 
\begin{equation}\label{eq:Erot}
	E_{{\rm rot}} = \frac{1}{2} I  \Omega^2
\,.
\end{equation}
If we assume a fixed moment of inertia, the release of rotational energy associated with a spindown $-d\Omega/dt \equiv - {\dot \Omega}$ is 
\begin{equation}
	L_{{\rm rot}} = I \Omega {\dot \Omega}
\, .
\label{eq:Lrot}
\end{equation}
For a magnetized star with a wind of mass loss rate ${\dot M}$, \citet[][]{wd1967} argued that the loss of the star's angular momentum $J = I \Omega$ scales as

\begin{equation}
	{\dot J} = I  {\dot \Omega} = {\dot M}  \Omega R_{\rm A}^2
\, ,
\label{eq:Jdot}
\end{equation}
which gives the associated release of rotational luminosity the scaling,

\begin{equation}
	L_{{\rm rot}} = {\dot M}  \Omega^2 R_{\rm A}^2 
\, .
\label{eq:Lrot}
\end{equation}

For a star with an equatorial field of strength $B_{\rm eq}$ at the stellar surface radius $R_\ast$, \cite{ud2002} and \cite{ud2008} showed that the Alfv\'en radius depends on the dimensionless wind magnetic confinement parameter $\eta_\ast$ (Eqn. \ref{eq:etastar}). Specifically, for a magnetic multipole of order $p$ (=1, 2 for monopole, dipole, etc.), with radial scaling as $B \sim r^{-(p+1)}$, \ra~scales as

\begin{equation}
\frac{R_{\rm A}}{R_\ast} =  \eta_\ast^{1/2p}
\, ,
\label{eq:RA}
\end{equation}
which for the standard dipole case ($p=2$), reduces to the scaling given in Eqn.\ (\ref{eq:RAdip}).

The \cite{wd1967} analysis treated the simple case of a pure radial field from a spilt monopole, with $p=1$. But \cite{ud2008} showed a base dipole field leads to a spindown that follows the \cite{wd1967} scaling (Eqn.\ \ref{eq:Jdot}), where $R_{\rm A}$ is given by Eqn. \ref{eq:RA} with a multipole index set to the $p=2$ value for a dipole.

\subsection{Breakout from centrifugal magnetospheres}

The above wind-confinement scalings work well for wind-magnetic braking, which operates through wind stress on open field lines, wherein the associated Poynting flux carries away most of the angular momentum.

But for rapid rotators with a strong field, the magnetic trapping of the wind into a centrifugal magnetosphere leads to some important differences for scalings of the associated luminosity.

First, as discussed in the appendices of \citet[][see their equation A7]{town2005c}, for trapping and breakout from a CM, the wind speed  $v_\infty$ in the usual wind confinement parameter $\eta_\ast$ is replaced with a characteristic dynamical speed of the stellar gravity, which we take here  to be the surface orbital speed $v_{{\rm orb}} \equiv \sqrt{G M_\ast/R_\ast}$ (since this is used in the definition of $W$ and thus $R_{\rm K}$), giving the {\em centrifugal} magnetic confinement parameter
\begin{equation}
	\eta_{\rm c}  \equiv \frac{B_{\rm d}^2 R_\ast^2}{{\dot M} v_{{\rm orb}}}
\,.
\label{eq:etac}
\end{equation}

A second difference stems from the fact that, even for an initially dipolar field, the rotational stress of material trapped in the CM has the effect of stretching the field outwards, thus weakening its radial drop off, and so reducing the effective multipole index to $p<2$.

Finally, this stretching ultimately leads to centrifugal breakout (CBO) events, with associated release of energy via magnetic reconnection. In general the overall total luminosity available from CBO events should follow a general scaling analogous to that for $L_{{\rm rot}}$,



\begin{equation}
\boxed{
L_{\rm CBO} \approx {\dot M} \Omega^2 R_\ast^2 \eta_{\rm c}^{1/p}
}
\, .
\label{eq:Lcbo}
\end{equation}

\subsubsection{Split monopole case}\label{subsubsec:split_monopole}

As a first example, consider the limit in which field lines are completely opened by the wind ram pressure into a split monopole, with $p=1$, which gives

\begin{eqnarray}
L_{\rm CBO}(p=1) &\equiv& {\dot M}  \Omega^2 R_\ast^2 \eta_{\rm c}
\nonumber
\\
&=& \frac{\Omega^2 R_\ast^4 B_{\rm d}^2}{v_{\rm orb}} 
\nonumber
\\
&=& W \Omega R_\ast^3 B_{\rm d}^2 
\, ,
\label{eq:L1}
\end{eqnarray}

\noindent where $W$ is the critical rotation fraction (Eqn.\ \ref{eq:wrot}). Note that the second equality recovers the empirical scaling $L_{\rm rad} \propto B^2R_\ast^4/P_{\rm rot}^2$ found by \cite{2021MNRAS.507.1979L} and verified in Paper I.

Remarkably, note also that in this monopole field case the dependence on wind feeding rate ${\dot M}$ has {\em canceled}. Dimensionally, the scaling now is as if the total magnetic energy over a volume set by $R_\ast^3$ is being tapped on a rotational timescale.

An alternative physical interpretation is that the field acts more like a conduit, trapping mass in a CM, with total rotational energy tapped on a breakout timescale, set in this monopole case by the orbital timescale.

\subsubsection{Dipole case}

More generally, this breakout luminosity depends on the wind feeding rate.

In particular, for the pure dipole scaling with $p=2$, we find

\begin{eqnarray}
L_{\rm CBO}(p=2) &=& {\dot M} \Omega^2 R_\ast^2 \eta_{\rm c}^{1/2} 
\nonumber
\\
&=& \frac{L_{\rm CBO}(p=1)}{\sqrt{\eta_{\rm c}}}
\, .
\label{eq:L2}
\end{eqnarray}

This has $L_{\rm CBO} \sim \sqrt{\dot M}$, with a weaker, linear scaling with $B_{\rm d}$.

In general, empirical evaluation of $L_{\rm CBO}$ thus requires evaluation of the wind feeding rate ${\dot M}$, where we have used the same CAK mass-loss rates as adopted in Paper I.

In the applications below, we consider multipole indices $1 < p < 2$, intermediate between these monopole and dipole limits.

\subsection{Application to radio emission}

   \begin{figure}
   \centering
   \includegraphics[width=0.5\textwidth]{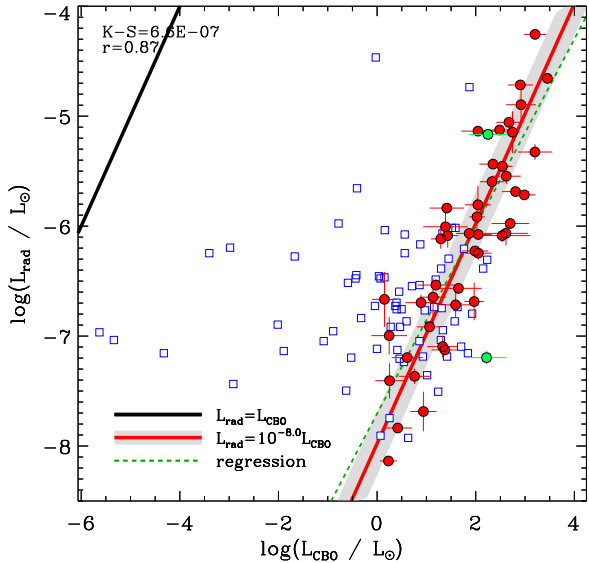} 
      \caption[]{Radio luminosity $L_{\rm rad}$ as a function of breakout luminosity $L_{\rm CBO}$ for the split monopole case. The thick black line indicates $L_{\rm rad}=L_{\rm CBO}$; the solid red line shows the same line shifted by the mean difference. The dashed green line shows the regression. Red circles indicate stars with detected emission; blue squares, stars without emission. Green circles indicate HD\,171247 and HD\,64740.}
         \label{lrad_lcbo}
   \end{figure}

Let us next consider how well such breakout scalings for rotational luminosity correlate with observed radio luminosities, $L_{\rm rad}$.
Noting that the dimensional scaling of breakout luminosity is set by the $p=1$ case, it is convenient to cast the general scaling in the form

\begin{equation}
L_{\rm CBO} (p) =  L_{\rm CBO}(p=1) \eta_{\rm c}^{-1+1/p} = L_{\rm CBO}(p=1) \eta_{\rm c} ^q
\, ,
\label{eq:Lp}
\end{equation}

\noindent by which we see an inferred empirical exponent $q$ in $\eta_{\rm c}$ implies an effective multipole exponent $p=1/(1+q)$.


Let us first examine how well this basic, dimensional scaling of the monopole model, with $p=1$ and so $q=0$, fits the observed radio emission.
Fig.\ \ref{lrad_lcbo} shows the observed radio luminosity $L_{\rm rad}$ as a function of $L_{\rm CBO}$ for the monopole case (Eqn.\ \ref{eq:L1}, i.e.\ with no dependence on \mdot). The thick black line shows $L_{\rm rad} = L_{\rm CBO}$, while the thick red line shows the same relationship shifted by about 8 dex, the mean difference between $L_{\rm rad}$ and $L_{\rm CBO}$ for radio-bright stars for the monopole scaling. This line is almost indistinguishable from a regression of $L_{\rm rad}$ vs.\ $L_{\rm CBO}$ for radio-bright stars. The relationship yields a correlation coefficient $r = 0.87$, and separates radio-bright from radio-dim stars with a K-S probability of about $10^{-8}$. Further, there are very few radio-dim stars with scaled breakout luminosities greater than the upper limits on their radio luminosities, i.e.\ to the right of the red line; those radio-dim stars that are to the right of the line, are very close to it.

   \begin{figure}
   \centering
   \includegraphics[width=0.5\textwidth]{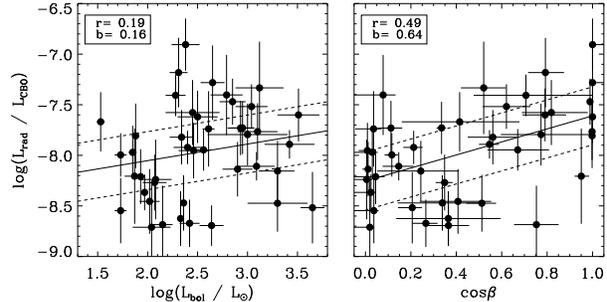}
      \caption[]{Ratio between $L_{\rm rad}$ and $L_{\rm CBO}$ as a function of bolometric luminosity ({\em left}) and $\cos{\beta}$ (right).}
         \label{lradlcbo_resid_lum_beta}
   \end{figure}

Eqn.\ \ref{eq:L1} does not yield quite as high of a correlation coefficient as the purely empirical scaling in Paper I. To see if there is some dependence on the mass-loss rate, the left panel of Fig.\ \ref{lradlcbo_resid_lum_beta} shows the residual radio luminosity after subtraction of the monopole $L_{\rm CBO}$ as a function of bolometric luminosity. There is only a weak dependence on $L_{\rm bol}$, with a correlation coefficient $r = 0.19$ and a slope $b = 0.16$. 

Another factor that may affect radio luminosity is the obliquity angle $\beta$ of the magnetic dipole axis from the rotational axis. Indeed, in the empirical regression analysis in Paper I the factor $f_\beta = (1 + \cos{\beta})/2$ was found to improve the correlation. The plasma distribution in the CM is a strong function of $\beta$, since the densest material accumulates at \rk~at the intersections of the magnetic and rotational equatorial planes \citep[][]{town2005c}. For the special case of an aligned rotator ($\beta = 0^\circ$) this will result in plasma being evenly distributed around \rk. With increasing $\beta$ the plasma distribution becomes increasingly concentrated at the two intersection points, leading to a warped disk that eventually becomes two distinct clouds. Therefore, the mass confined within the CM will be a maximum for $\beta = 0^\circ$ and a minimum for $\beta = 90^\circ$. If reconnection in the CM is the source of the high-energy electrons that populate the radio magnetosphere, we would then naturally expect that radio luminosity should decrease with increasing $\beta$. The right panel of Fig.\ \ref{lradlcbo_resid_lum_beta} shows the residual radio luminosity as a function of $\cos{\beta}$, and demonstrates that radio luminosity in fact does increase with decreasing $\beta$; in fact, the relationship is much stronger than for $\log{L_{\rm bol}}$, with $r = 0.49$ and $b = 0.64$. 

   \begin{figure}
   \centering
   \includegraphics[width=0.5\textwidth]{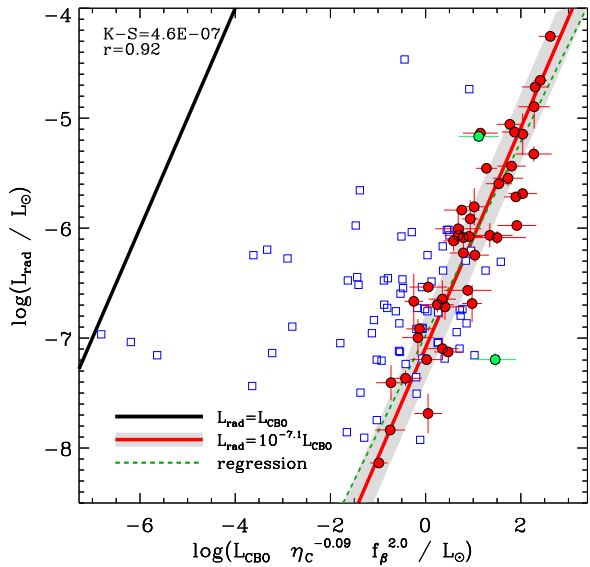} 
      \caption[]{As Fig.\ \ref{lrad_lcbo}, with $L_{\rm CBO}$ modified to account for the residual dependence on bolometric luminosity (i.e.\ the surface mass-flux from the wind) and the tilt angle of the magnetic dipole (Fig.\ \ref{lradlcbo_resid_lum_beta}).}
         \label{fbeta_etc_lrad_lcbo}
   \end{figure}

Fig.\ \ref{fbeta_etc_lrad_lcbo} replicates Fig.\ \ref{lrad_lcbo}, with the difference that corrections for \mdot~and $\beta$ are accounted for. Following Eqn.\ \ref{eq:Lp}, \mdot~dependence was determined by scaling Eqn.\ \ref{eq:L1} with $\eta_{\rm c}^q$. A purely empirical correction for $\beta$ was adopted as $f_{\beta}^x = ((1 + \cos{\beta})/2)^x$, such that $f_\beta(\beta = 0^\circ) = 1$ and $f_\beta(\beta = 90^\circ) \neq 0$. By minimizing the residuals, the best-fit exponents are $q = -0.09$ and $x = 2$. The former exponent corresponds to $p = 1.1$, implying only a very slight departure from the monopole scaling. The latter indicates an increase in $L_{\rm rad}$ by a factor of 4 as $\beta$ decreases from $90^\circ$ to $0^\circ$. As can be seen in Fig.\ \ref{fbeta_etc_lrad_lcbo}, these corrections lead to a tighter correlation ($r = 0.92$) and a somewhat reduced ratio between $L_{\rm CBO}$ and $L_{\rm rad}$ to around 7 dex. 


\subsubsection{Emission threshold}

   \begin{figure}
   \centering
   \includegraphics[width=0.5\textwidth]{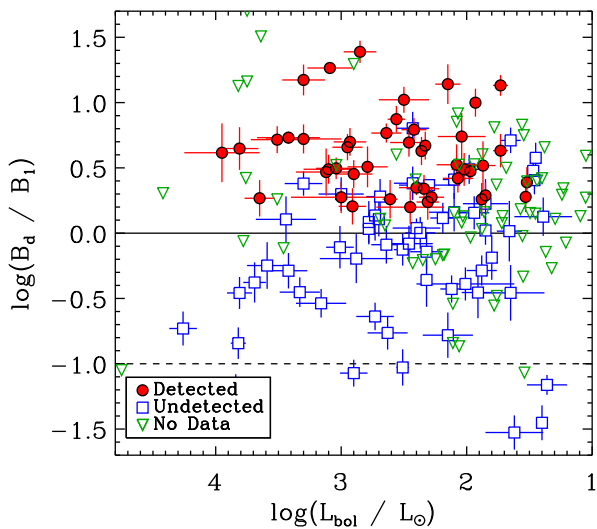} 
      \caption[]{Ratio between surface magnetic field strength and the threshold magnetic field strength necessary to achieve a flux density of 1 mJy, as a function of bolometric luminosity.}
         \label{bthresh}
   \end{figure}

In their development of a breakout scaling relationship for H$\alpha$ emission from CMs, \cite{2020MNRAS.499.5366O} defined a threshold magnetic field strength $B_{\rm K1}$ as the strength of the magnetic field at \rk~necessary to confine a sufficient quantity of plasma at \rk~for the optical depth to reach unity. They demonstrated that all magnetic early B-type stars with $B_{\rm K} / B_{\rm K1} > 1$ are H$\alpha$-bright, while all stars with $B_{\rm K} < B_{\rm K1}$ do not display H$\alpha$ emission. By solving Eqn.\ \ref{eq:L1} for $B_{\rm d}$ we can derive a similar threshold value for the radio luminosity:

\begin{equation}\label{eqn:bthresh}
B_{\rm thresh} = \left(\frac{\epsilon L_{\rm CBO} P_{\rm rot}}{2 \pi R_\ast^3 W}\right)^{1/2},
\end{equation}

\noindent where $\epsilon \sim 10^{-8}$ is an efficiency scaling determined from the empirical ratio between $L_{\rm CBO}$ and $L_{\rm rad}$, and additional dependence on $\beta$ and $\eta_{\rm c}$ is implic itly ignored. Since what is actually observed is a flux density $F_{\rm rad}$ rather than a luminosity $L_{\rm rad} \propto F_{\rm rad} d^2$, Eqn.\ \ref{eqn:bthresh} is necessarily a function of distance $d$. Amongst the radio-bright stars, the median flux density uncertainty is 0.1 mJy, while the median significance of a detection is around 10$\sigma$ i.e.\ 1 mJy. We therefore take the flux density detectability threshold for the sample as 1 mJy and solve Eqn. \ref{eqn:bthresh} accordingly to obtain $B_1$ (i.e. the surface magnetic field necessary to generate 1 mJy of flux density at the star's distance). The results are shown in Fig.\ \ref{bthresh}. 

As expected, all radio-bright stars have $\log{B_{\rm d} / B_1} \gtrsim 0$. The relationship for non-detected stars is not as clean as for the similar plot for H$\alpha$ shown by \cite{2020MNRAS.499.5366O}, as there are a large number of stars with surface magnetic fields above this threshold. However, the radio observations comprising this sample, having been obtained at a variety of observatories with different capabilities over a span of over 30 years, are quite heterogeneous, with a wide range of upper limits, and many of the non-detected stars in this regime have upper limits comparable to 1 mJy. Further, generally only a single snapshot at one frequency is available, and it is possible that they were observed at inopportune rotational phases. These stars should certainly be reobserved with modern facilities. 

The dashed line in Fig.\ \ref{bthresh} shows the theoretical detection limit for radio telescopes such as the upcoming Square Kilometre Array able to achieve $\mu$Jy precision, under the assumption that a 10$\sigma$ detection (i.e.\ 10 $\mu$Jy) is necessary for the star's radio emission to be securely detected. As can be seen, such facilities can at least double the number of stars with measured gyrosynchrotron emission. This is especially true when stars that have not yet been observed in the radio are included: the green triangles in Fig.\ \ref{bthresh} show those stars from the samples studied by \cite{2007AA...475.1053A}, \cite{2019MNRAS.483.2300S,2019MNRAS.483.3127S}, and \cite{2019MNRAS.490..274S} without radio observations, essentially all of which are expected to have radio flux densities es above 10 $\mu$Jy and about half of which should have flux densites above 1 mJy. 

\subsubsection{How CBO reconnection can  lead to radio emission}

Breakout events are accompanied by centrifugally driven reconnection of magnetic fields that have been stretched outward by rotational stress acting against the magnetic tension of the initially closed loops. As these loops reconnect, the associated release of magnetic energy can strongly heat the ejected plasma. Some fraction of this reconnection energy can accelerate both ions and electrons to highly super-thermal energies, with some of these particles becoming trapped into gyration along closed magnetic field loops near the reconnection site. The associated gyrosynchrotron emission of the much lighter electrons can then produce the observed radio emission. 

The basic scenario of electron acceleration in reconnection events, followed by gyrosynchrotron emission along magnetic loops, is indeed already a central component of the model for radio emission \citep{2004A&A...418..593T}. However, this model is based on {\em wind}-driven reconnection, with no inclusion for any role of stellar rotation. As such, the available reconnection luminosity is expected to scale with the wind kinetic energy $L_{\rm wind} = {\dot M} v_\infty^2/2$. By comparison, the rotational luminosity for a multipole exponent $p$ is larger by a factor

\begin{eqnarray}
\frac{L_{\rm CBO}(p)}{L_{\rm wind}} 
&=& 2 \eta_{\rm c}^{1/p} \,  \left ( \frac{v_{\rm rot}}{v_\infty} \right )^2
\nonumber
\\
&=& \eta_{\rm c}^{1/p} \, W^2 \left ( \frac{v_{\rm esc}}{v_\infty} \right )^2
\nonumber
\\
&\approx& \frac{\eta_{\rm c}^{1/p} \, W^2}{9}
\, ,
\label{eq:LpbLw}
\end{eqnarray}

\noindent where the last equality stems from the standard result that the stellar wind speed scales with the escape speed as $v_\infty \approx 3 v_{\rm esc}$. For typical values for B-star magnetospheres with $\eta_{\rm c} \approx 10^6$, $W \approx 1/2$  \citep{petit2013} and $p=4/3$, we find $L_{\rm CBO}(p=4/3)/L_{\rm wind} \approx  880$ (using the empirically derived value of $p=1.1$ yields an even greater ratio of almost 8000). For these $W$ and $p$ values, the ratio is greater than unity for even moderate confinement values $\eta_{\rm c} > 120 $.

Regardless of the relative values, a central empirical result here is the finding that $L_{\rm rad}$ has a clear scaling with rotation frequency as $\Omega^2$ and with surface field as $B_{\rm d}^{2/p}$, dependences which are entirely missing from $L_{\rm wind}$. Indeed, we find that $\log(L_{\rm rad})$ shows only weak correlation with $L_{\rm wind}$, with $r=0.3$. This strongly disfavors the wind-driven reconnection model proposed by \cite{2004A&A...418..593T}; 
but it is consistent with the scenario proposed here
that centrifugal-breakout reconnection provides the underlying energy that leads to the radio luminosity through gyrosynchrotron emission.

While we have cast available energy in terms of loss of the star's rotational energy, one has to be careful not to take this too literally. Most (70+\%) of the angular momentum loss in spindown is through magnetic field Poynting stresses. But the CBO material that leads to reconnection should share the same basic Weber-Davis scaling with $R_{\rm A}$, and it is that component that this scenario associates with the reconnection and the resulting electron acceleration and radio emission.

\section{Discussion}\label{sec:discussion}
\label{sec:disc}

\subsection{Comparison with alternative theoretical interpretations}

The empirical scaling relationship discovered by \cite{2021MNRAS.507.1979L} and confirmed in paper I, $L_{\rm rad} \propto B_{\rm d}^2R_\ast^4/P_{\rm rot}^2 = (\Phi/P_{\rm rot})^2$, is explained above as a consequence of electron acceleration via centrifugal breakout. However, \citeauthor{2021MNRAS.507.1979L} pointed out that $\Phi/P_{\rm rot}$ has the physical dimension of an electromotive force $\emf$ (EMF), which they speculated may be suggestive of an underlying theoretical mechanism. In this section we examine gyrosynchrotron emission from this standpoint.
We perform a theoretical analysis to test if the physical conditions able to sustain large scale electric currents within the stellar magnetosphere can be verified.

This empirical association by \cite{2021MNRAS.507.1979L} of radio emission with the voltage of an EMF also stands in contrast with the previous theoretical model by \cite{2004A&A...418..593T}, which associates the acceleration of radio-emitting non-thermal electrons with the wind-induced current sheet that forms in the middle magnetosphere. However, \cite{2021MNRAS.507.1979L} conclusively demonstrated that the wind does not provide sufficient power to the middle magnetosphere to drive the observed levels of radio emission.

The highly ionized plasma in these magnetospheres implies a very high conductivity, and so currents can form even with a vanishingly small EMF. Instead, the current density $J$ is set by Ampere's law as a result of a curl induced in a stressed field,
\beq
J = \frac{c}{4 \pi} \, \nabla \times B \, .
\eeq
Even in a non-rotating wind-fed magnetosphere, large-scale stressing of the magnetic field by the wind ram pressure forces outlying closed loops to open,  with a Y-type neutral point at the top of the last closed loop; above this there develops a split monopole, with a current sheet separating field lines of opposite polarity. But unless there are instabilities or induced variability, this current sheet does not by itself lead to energy dissipation that can heat the plasma or accelerate electrons. This, together with the lack of observed radio emission from stars with slow rotation, thus strongly disfavors the \cite{2004A&A...418..593T} model based on the wind-induced current sheet.

While the  \cite{2021MNRAS.507.1979L} {\em empirical} association of observed radio emission with an EMF is interesting and insightful,
there are some challenges to using this as a basis for a self-consistent {\em theoretical} model.
The general principles regarding current vs.\ EMF scenarios can be well illustrated by a simple circuit model, using an Ohm's law $I=\emf/\res$ to related current $I$ and EMF $\emf$ through a resistance $\res$. The associated dissipated power, or luminosity, scales as
\beq
L_{\rm emf} = I \emf 
= I^2{\res}
= \frac{\emf^2}{\res} 
\, .
\label{eq:Lemf1}
\eeq
If one fixes the current $I$ (as induced by the globally imposed magnetic curl), then the second equality shows that in the limit of vanishing resistivity $\res \rightarrow 0$, the luminosity also vanishes; this underlies one fundamental issue with the current-sheet model advocated by \cite{2004A&A...418..593T}.

To understand the ramifications of the last equality of Eqns. \ref{eq:Lemf1} within the context of the empirical association of the gyrosynchrotron scaling law with an EMF, let us return to consideration of plasma conditions in such magnetospheres. In the notation of the present paper, the EMF can be written as $\emf = B_\ast R_\ast^2 \Omega/c$, where the speed of light $c$ comes in from the CGS form for the induction equation. In terms of a plasma resistivity $\rho$ (with units of time), the circuit resistance scales with resistivity times a length over area, which in this context gives $\res \approx \rho R_\ast/R_\ast^2 = \rho/R_\ast$. Thus Eqn.\ \ref{eq:Lemf1} becomes

\beqa
L_{\rm emf} &=& \frac{B_\ast^2 R_\ast^5 \Omega^2}{\rho c^2} 
\label{eq:Lemf2}
\\
&=& (B_\ast^2 R_\ast^3 \Omega ) \, \left [ \frac{ v_{\rm rot} R_\ast}{\rho c^2} \right ]
\, ,
\label{eq:Lemf3}
\eeqa

\noindent where in the latter equality, $v_{\rm rot} = \Omega R_\ast$ is the surface rotation speed at the stellar equator. Here the term in parenthesis separates out the dimensional luminosity, while the dimensionless ratio in square brackets can be identified as a {\em magneto-rotational Reynold's number}, 

\beq
Re_{\rm mr} \equiv \frac{ v_{\rm rot} R_\ast}{\rho c^2} 
\, .
\label{eq:Rem}
\eeq

Rather remarkably, Eqn. \ref{eq:Lemf3} has a form very similar to that derived above (cf. Eqn. \ref{eq:L1}) for the monopole ($p=1$) CBO model. However an important, indeed crucial difference is that the CBO rotation scaling is relative to the near-surface orbital speed, $v_{\rm orb} = \sqrt{GM_\ast/R_\ast}$. As such in this CBO model the associated dimensional rotation parameter $W = v_{\rm rot}/v_{\rm orb}$ is always {\em less than unity}.

Indeed, eqn. (\ref{eq:Lemf2}) is very similar to the  scaling invoked by
\citet[][see their Eqn. 4]{2001JGR...106.8101H} 
to model auroral emission from Jupiter.
In this case, the magnetospheric EMF accelerates ions and electrons to high-energy, which upon penetrating into the underlying Jovian atmosphere is dissipated through the low atmospheric conductance $\Sigma_{\rm J}$ (corresponding to high resistivity $\rho$), resulting in heating and associated thermal bremstrahlung to give auroral emission.

In contrast, because of the typically very low resistivity of the ionized plasma in hot-star magnetospheres, the associated dimensionless Reynolds number is expected to be very large. 
The associated dissipation luminosity (Eqn.\ \ref{eq:Lemf3}) in this EMF scenario would thus be enormous, leading in effect to a ``short circuit" that would quickly draw down the available pool of magnetic energy.

In principle, a theoretical model grounded in the EMF could invoke a stronger resistivity in some local dissipation layer, which would enter directly into the predicted scalings for the generated luminosity.
But it is unclear how this small-scale dissipation could be reconciled with the large-scale EMF that is taken to scale with the stellar radius,
and how such a dissipation could remain fixed over the range of stellar and magnetospheric parameters, in order to preserve the 
inferred empirical scaling of the observed radio luminosity with the global EMF.

These difficulties with an association of gyrosynchrotron scaling with EMF, and with the current sheet model, stand in contrast to some key advantages of the CBO mechanism proposed here.

First, this CBO model specifies a more modest magnetic dissipation rate, set by the base dimensional rate $B_\ast^2 R_\ast^3 \Omega$ {\em reduced} by the rotation factor $W<1$, instead of the enormous $Re_{\rm rm} \sim 10^{12}$ enhancement of an EMF mechanism. This CBO dissipation can be quite readily replenished over time by the centrifugal stretching of closed magnetic field lines by the constant addition of mass from the stellar wind. As such, the ultimate source of energy thus comes not from the field -- which acts merely as a conduit -- but from the star's rotational energy.

Second, these eventual centrifugal breakout events lead naturally and inevitably to magnetic reconnection. This thus preserves the longstanding notion \citep{2004A&A...418..593T} that such reconnection provides the basic mechanism to accelerate electrons to high energies, whereupon the gyration along the remaining field lines connecting back to the star results in the gyrosynchrotron emission of the observed radio luminosity.
 
Third, and perhaps most significantly, instead of the previous notion \citep{2004A&A...418..593T} that this reconnection is driven by the stellar wind -- with no consideration of any role for stellar rotation -- our model for CBO-driven reconnection puts rotation at the heart of the process, and so yields a scaling for luminosity that matches the strong dependence on rotation rate, as well as on magnetic field energy. Indeed, while the wind can certainly open the magnetic field and lead to the formation of a current sheet, this does not itself provide a power source, but merely results in a slower radial decline of the magnetic field strength as compared to that of a dipole. By contrast, CBO provides a clear power source for the acceleration of electrons to high energies. 

Thus, although only a small fraction of the breakout luminosity $L_{\rm CBO}$ ends up as radio luminosity, with an inferred effective efficiency $\epsilon \approx 10^{-8}$, the strong correlation between observed and predicted scalings provides strong empirical support for such a CBO model.

\subsection{Energy source - magnetic or rotational?}

The energy term in Eqn.\ \ref{eq:L1} is $B^2R^3$, and it would therefore be natural to assume that the magnetic field is the energy source powering radio emission. However, as suggested in \S~\ref{subsubsec:split_monopole}, this is probably not the case. The mean magnetic energy $E_{\rm mag}$ amongst the radio-bright stars is about $10^{42}$ erg, whereas the mean rotational kinetic energy $E_{\rm rot}$ in the same sub-sample is about $10^{47}$ erg, i.e.\ the star's rotation is a vastly greater energy reservoir. Indeed $E_{\rm mag} > E_{\rm rot}$ for only 3 stars (HD\,46328, HD\,165474, and HD\,187474), all of which have $P_{\rm rot} \sim$~years (and none of which are, of course, detected at radio frequencies). 

An additional consideration is that, if the magnetic field were the energy source, radio emission should over time draw down the magnetic energy of the star. The peak radio luminosity is around $10^{29}~{\rm erg~s^{-1}}$, implying that the magnetic energy of the most radio-luminous stars would be consumed in about $10^{13}~{\rm s} \sim 0.3~{\rm kyr}$. To the contrary, fossil magnetic fields are stable throughout a star's main sequence lifetime. For Ap/Bp stars below about 4\,M$_\odot$, the decline in surface magnetic field strength is entirely consistent with flux conservation in an expanding stellar atmosphere \citep[e.g.][]{2006A&A...450..763K,2007A&A...470..685L,2019MNRAS.483.3127S}, while for more massive stars there is an additional, gradual decay of flux \citep[e.g.][]{2007A&A...470..685L,2008A&A...481..465L,2016A&A...592A..84F,2019MNRAS.490..274S} that is however, much longer than the abrupt field decay timescale that would be implied if the breakout luminosity was powered by the magnetic field. Furthermore, the most plausible mechanism for flux decay is found in small-scale convective dynamos formed in the opacity-bump He and Fe convection zones inside the radiative envelope \citep[e.g.][]{2019MNRAS.487.3904M,2020ApJ...900..113J}, which naturally explains why flux does not decay in A-type stars (which lack these convection zones), and why flux apparently decays more slowly for the strongest magnetic fields \citep{2019MNRAS.490..274S} since strong fields inhibit convection \citep{2013MNRAS.433.2497S,2019MNRAS.487.3904M}. 

In contrast to the magnetic field, which decays slowly or not at all, magnetic braking is quite abrupt \citep{2019MNRAS.490..274S,2020MNRAS.493..518K}, making the larger rotational energy reservoirs of rapidly rotating stars a more far more plausible power source. Quantitatively, for the most radio-luminous stars in the sample it would take about 30 Myr for the energy radiated by gyrosynchrotron emission to remove the total rotational energy of the star. For stars with masses above 5\,M$_\odot$ (the mass range of the brightest radio emitters), this is comparable to or greater than the main sequence lifetime.

It therefore seems that the magnetic field cannot serve as the energy source, but rather acts as a conduit for the extraction of rotational energy and its conversion into gyrosynchrotron emission. The magnetic energy lost in breakout events is immediately replenished as mass is injected into the CM by the wind, with the ion-loaded magnetic field then stretching under the centrifugal stress acting on the co-rotating plasma.

\subsection{The case of Jupiter}

\cite{2021MNRAS.507.1979L} showed that the scaling relationship for the non-thermal radio emission from dipole-like rotating magnetospheres also fits the radio luminosity of Jupiter\footnote{This radio emission arises within Jovian magnetosphere, and so is distinct from the optical auroral emission discussed above, which arises from interactions in the upper Jovian atmosphere.}, suggesting an underlying similarity in the physics driving gyrosynchrotron emission from giant planets and magnetic hot stars. Adopting the same parameters as used by \citeauthor{2021MNRAS.507.1979L} ($B_{\rm eq} = 4$~G, $P_{\rm rot} = 0.41$~d, $M_{\rm J} = 1.9 \times 10^{27}$~kg, and $R_{\rm J} = 7.1 \times 10^5$~km) gives $W = 0.3$. The breakout luminosity is then $\log{L_{\rm CBO} / L_\odot} \sim -15.9$ or, at 1 cm, $L_\nu \sim 10^7 {\rm erg~s^{-1}~Hz^{-1}}$, translating to an expected flux {\em density} of around 40 Jy at a distance of 4 AU. This is about an order magnitude higher than the observed radio luminosity of Jupiter \citep{2003Icar..163..449D,2003Icar..163..434D}. However, it is worth noting that in the extrapolation shown by \citeauthor{2021MNRAS.507.1979L}, Jupiter's EMF of 376 MV is near the lower envelope of the range of uncertainty inferred from hot stars, i.e.\ Jupiter is somewhat less luminous than predicted by a direct extrapolation of the hot star scaling relationship. Furthermore, 1 dex is at the upper range of the scatter about the $L_{\rm CBO}$ relationship (see Figs.\ \ref{lrad_lcbo} and \ref{fbeta_etc_lrad_lcbo}).

One possible explanation for Jupiter being less luminous than predicted is that Jupiter's primary ion source, the volcanic moon Io, is effectively a point source offset from the centre of the Jovian magnetosphere. This is in contrast to stellar winds, which feed the magnetosphere isotropically and continuously from the centre. The result is that hot star magnetospheres are relatively more populated, and there is therefore more material available for the generation of gyrosynchrotron emission. Another potential issue is that in the Jovian magnetosphere reconnection takes place in the magnetotail due to stretching by the solar wind; its azimuthal extent will therefore be limited, in analogy to the obliquity dependence found in stellar magnetospheres. Exploring whether the approximate consistency between the Jovian and stellar radio luminosities is indeed due to a similarity in the underlying physics, or is merely coincidental, will require a detailed analysis that is outside the scope of this paper. 

\subsection{A solution to the low-luminosity problem?}


It is notable that magnetospheres are detectable in radio frequencies in stars with CMs that are too small to be detectable in H$\alpha$. In addition to being a more sensitive magnetospheric diagnostic, this may also suggest an answer to the low-luminosity problem identified by \cite{2020MNRAS.499.5379S} and \cite{2020MNRAS.499.5366O}. While CBO matches all of the characteristics of H$\alpha$ emission from CM host stars, emission disappears entirely for stars with luminosities below about $\log{L_{\rm bol}/L_\odot} \sim 2.8$. This could be either a consequence of a `leakage' mechanism, operating in conjunction with CBO to remove plasma via diffusion and/or drift across magnetic field lines \citep{2018MNRAS.474.3090O}, or due to the winds of low-luminosity stars switching into a runaway metallic wind regime \citep{1992A&A...262..515S,1995A&A...301..823B,2002ApJ...568..965O}. In the former case the leakage mechanism only becomes significant when \mdot~is low. In the latter case, H$\alpha$ emission is not produced for the simple reason that the wind does not contain H ions. Notably, the peculiar surface abundances of magnetic stars may lead to enhanced mass-loss rates as compared to non-magnetic, chemically normal stars \citep{krticka2014}. 

Since CBO apparently governs gyrosynchrotron emission, and is seen in stars down to $\log{L_{\rm bol}/L_\odot} \sim 1.5$, the leakage scenario seems to be ruled out as an explanation for the absence of H$\alpha$ emission. This therefore points instead to runaway metallic winds. One possible complication is that, as is apparent from the direct comparison of H$\alpha$ emission equivalent widths to radio luminosities (see Fig.\ 6 in Paper I), stars without H$\alpha$ so far are also relatively dim in the radio (at least for those stars for which H$\alpha$ measurements have been obtained). These stars have systematically lower values of $B_{\rm K}$ than have been found in more luminous H$\alpha$-bright stars (Fig.\ 3 in Paper I). Thus, a crucial test will be examination of both H$\alpha$ and radio for a star with a luminosity well below 2.8, but $B_{\rm K} \sim 3$, i.e.\ it must be cool, very rapidly rotating ($P_{\rm rot} \sim 0.5$~d), and strongly magnetic $(B_{\rm d} \sim 10~{\rm kG})$. So far no such stars are apparently known. 

A further complication to the runaway wind hypothesis is provided by 36\,Lyn, a relatively cool ($T_{\rm eff} \sim 13$~kK), radio-bright star which, while it does not show H$\alpha$ emission, does display eclipses in H$\alpha$ \citep{2006AA...458..581S} and therefore must have H inside its magnetosphere which, presumably, originated in the stellar wind. Why no other star in 36\,Lyn's \teff~range should show evidence of a similar phenomenon is not currently understood, although its peculiar magnetosphere may be related to the remarkably high toroidal component of its magnetic field in comparison to other magnetic stars, in which the toroidal component is generally quite weak \citep{2018MNRAS.473.3367O,2019A&A...621A..47K}. Alternatively, this may be due to a simple selection effect: 36\,Lyn's eclipses are only detectable for about 10\% of its rotational cycle, and eclipse absorption would be broader and shallower for more rapidly rotating stars.

\subsection{X-rays from CBO?}
\label{sec:xray}

\begin{figure}
\begin{center}
\includegraphics[scale=0.35]{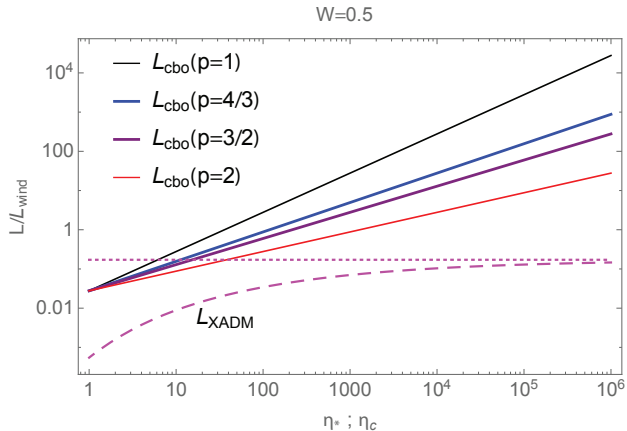}
\caption{For the standard case of half-critical rotation ($W=0.5$), comparison of the X-ray luminosities from  the XADM  model (dashed) 
of \citet{ud2014} with CBO models of various indices $p$ (solid), plotted vs.\ associated magnetic confinement parameter, with all luminosities normalized  by the kinetic energy luminosity of the stellar wind, $L_{\rm wind} = {\dot M} v_\infty^2/2$.
The horizontal dotted line represents the asymptotic luminosity for XADM in the strong-confinement limit.
Note that, for this $W=0.5$ case, the CBO luminosities are significantly enhanced over that from the XADM model.
Results for other rotations can be readily determined by scaling the CBO models with $W^2$. 
}
\label{fig:fig2}
\end{center}
\end{figure}

The reconnection energy from CBO events might also be an important for the X-rays observed from magnetic stars.

For slow rotators with only a dynamical magnetosphere, and no centrifugal magnetosphere component, the observed X-rays follow quite closely the scaling predicted by the ``X-rays from Analytic Dynamical Magnetosphere'' (XADM) model developed by 
\citet{ud2014}, as shown in Figure 6 of \citet{2014ApJS..215...10N}. 
With a 10\% scaling adjustment to account for the X-ray emission duty cycle seen in MHD simulations, the overall agreement between observed and predicted X-ray luminosities is quite remarkable for such DM stars (denoted by open circles and triangles), spanning more than four orders of magnitude in X-ray luminosity!

However there are several stars with observed X-ray luminosities well above (by 1-2 orders of magnitude) the 10\%-XADM scaling;
all are  CM stars.
These are the very stars that the analysis here predicts to have CBO reconnection events that could power extra X-ray emission, and so supplant the X-rays from wind confinement shocks predicted in the XADM analysis.

To lay a basis to examine whether CBO reconnection X-rays might explain this observed X-ray excess for CM stars, Figure \ref{fig:fig2} compares the CBO vs. XADM predicted scalings for X-ray luminosity, both normalized by the wind kinetic energy luminosity $L_{\rm wind} = {\dot M} V_\infty^2/2$, and plotted versus their associated confinement parameter $\eta_{c}$ or $\eta_\ast$.
The XADM plot is based on eqn. (42)  of \citet{ud2014}, with velocity exponent $\beta =1$.
As seen from eqn.\ (\ref{eq:LpbLw}) here, the CBO scalings also depend on the multipole exponent $p$, so the various curves show results for various $p$, as given by the legend.
Also, the values shown are for a {\em fixed}, fiducial value for the critical rotation fraction $W=1/2$, but the results for other rotations can be readily determined by 
scaling with $W^2$. 

Note that the CBO scalings increase as $\eta_{c}^{1/p}$, while the XADM scaling saturates at large $\eta_\ast$.
For the chosen default rotation $W=0.5$, the CBO scalings are generally above those for the XADM, but this will change for lower $W$.

To test this possibility that CBO plays a role in augmenting X-ray emission, a next step should be to test whether the observed X-rays from CM stars follow the CBO scaling with $\eta^{1/p} W^2$, as given by eqn. (\ref{eq:LpbLw}) when scaled to $L_{\rm wind}$, or more generally by eqns. (\ref{eq:etac}) and (\ref{eq:Lcbo}).


\section{Summary and Future Outlook}
\label{sec:sum}

The radio luminosities of the early-type magnetic stars were empirically found to be related to the stellar magnetic flux rate 
\citep[][Paper I]{2021MNRAS.507.1979L}.
\citet{2021MNRAS.507.1979L} did not provide a definitive physical explanation regarding the origin of the non-thermal electrons.
To provide the theoretical support for explaining how non-thermal electrons originate, in this paper
we have extended the centrifugal breakout model that successfully predicts the H$\alpha$ emission properties of stars with centrifugal magnetospheres \citep{2020MNRAS.499.5379S,2020MNRAS.499.5366O}, deriving a breakout luminosity $L_{\rm CBO} \propto (B^2R_\ast^3/P_{\rm rot})W$, where the first term in brackets has natural units of luminosity, and the dimensionless critical rotation parameter $W$ is an order-unity correction that includes the additional $R_\ast$ and $P_{\rm rot}$ dependence. The radio luminosity is then $L_{\rm rad} = \epsilon L_{\rm CBO}$, where $\epsilon \sim 10^{-8}$ is an efficiency factor. Crucially, there is a nearly 1:1 correspondence between $L_{\rm rad}$ and $\epsilon L_{\rm CBO}$. 

The basic scaling relationship is appropriate for a split monopole. Generalization to higher-order multipoles is accomplished with a correction $\eta_{\rm c}^{1/p}$, where $\eta_{c}$ is the centrifugal magnetic confinement parameter and $p$ is the multipolar order (1 for a monopole, 2 for a dipole, etc.). The small residual dependence of radio luminosity on bolometric luminosity is removed by adopting $p \sim 1.1$, i.e.\ a nearly monopolar field. The minimal residual dependence on $L_{\rm bol}$ (which in line-driven wind theory sets the mass loss rate through a scaling ${\dot M} \sim L_{\rm bol}^{1.6}$) confirms that the radio magnetosphere is nearly independent of the mass-loss rate. However, we find that there is a stronger dependence of the residuals on the obliquity $\beta$ of the magnetic axis with respect to the rotation axis, with $L_{\rm rad}$ increasing by about a factor of 4 from $\beta = 90^\circ$ to $0^\circ$. This is consistent with expectations from the rigidly rotating magnetosphere model that the amount of plasma trapped in a centrifugal magnetosphere is a strong function of $\beta$ \citep{town2005c}, since with less plasma in the CM, there will be fewer electrons available to populate the radio magnetosphere.

While radio emission and H$\alpha$ emission are explained by a unifying mechanism, they probe different parts of the magnetosphere as well as different parts of the centrifugal breakout process. H$\alpha$ emission probes the cool plasma trapped in the CM, which has not yet been removed by breakout. During a breakout event, some of the energy released by magnetic reconnection accelerates electrons to relativistic velocities, which then return to the star, emitting gyrosynchrotron radiation as they spiral around magnetic field lines. Following the result reported in this paper, we explain the radiation belt model proposed by \citet{2021MNRAS.507.1979L} to be the magnetic shell connected to the centrifugal breakout region close to the magnetic equator. This largely preserves the \cite{2004A&A...418..593T} model, with the primary difference being the mechanism of electron acceleration.

Overall, the results here provide a revised foundation on which to build a detailed theoretical model for how centrifugal-breakout reconnection leads to acceleration of electrons and the associated radio gyrosynchrotron emission. In particular, we might be able to quantify the level of reconnection heating through MHD simulations, and how it scales with $W$, $\eta_\ast$, etc., as has been done for other scalings like spindown. 

Future theoretical work should focus on the details of the acceleration of the electrons through reconnection, and their subsequent gyrosynchrotron emission of polarized radio emission (and perhaps other observable spectral bands like X-rays), with the specific aim to understand, and quantitatively reproduce, the inferred emission efficiencies $\epsilon$.
This work should also extend to explore the connection with electron cyclotron maser (ECM) radio emission that has been detected in many of the same stars showing gyrosynchrotron emission 
\citep{2021arXiv210904043D}.

\section*{Acknowledgments}
MES acknowledges the financial support provided by the Annie Jump Cannon Fellowship, supported by the University
of Delaware and endowed by the Mount Cuba Astronomical Observatory. 
AuD acknowledges support by NASA through Chandra Award 26 number TM1-22001B issued by the Chandra X-ray Observatory 27 Center, which is operated by the Smithsonian Astrophysical Observatory for and on behalf of NASA under contract NAS8-03060. 
PC and BD acknowledge support of the Department of Atomic Energy, Government of India, under project no. 12-R\&D-TFR-5.02-0700.

\section*{Data Availability Statement}

 No new data were generated or analysed in support of this research. All data referenced were part of the associated Paper I, \citet{shultz2021}.

\bibliography{bib_dat.bib}{}
\label{lastpage}

\end{document}